\documentclass[dvipsnames]{llncs}
\IfFileExists{pdfsync.sty}{\usepackage{pdfsync}}{}

\usepackage{amsmath}
\usepackage{aeguill}
\usepackage{amsfonts}
\usepackage{amssymb}
\usepackage[pdftex,usenames,dvipsnames]{color}
\usepackage{graphicx}
\usepackage{here}

\newcommand\R{\mathbb{R}}
\newcommand\N{\mathbb{N}}
\newcommand\RP{\mathbb{R^+}}

\newcommand\vu{\vec{u}}

\begin{document}

\newcommand\LOCALORNOTLOCAL[2]{#2}
\providecommand\url[1]{\textit{#1}}


\newcommand\findocument{
  \bibliographystyle{apalike}
  \LOCALORNOTLOCAL{
   \bibliography{/Users/bournez/bibliographie/Bib-Files/Analog,/Users/bournez/bibliographie/Bib-Files/BSSComplexite,/Users/bournez/bibliographie/Bib-Files/BSSfunc,/Users/bournez/bibliographie/Bib-Files/perso,/Users/bournez/bibliographie/Bib-Files/hybridsystems,/Users/bournez/bibliographie/Bib-Files/systemesdynamiques,/Users/bournez/bibliographie/Bib-Files/HSIII,/Users/bournez/bibliographie/Bib-Files/HSI,/Users/bournez/bibliographie/Bib-Files/HSII,/Users/bournez/bibliographie/Bib-Files/TheorieDesJeuxCommunications,/Users/bournez/bibliographie/Bib-Files/LuEns,/Users/bournez/bibliographie/Bib-Files/complexite,/Users/bournez/bibliographie/Bib-Files/copieperso,/Users/bournez/bibliographie/Bib-Files/HSCC00,/Users/bournez/bibliographie/Bib-Files/HART97}
}
    {\bibliography{SurveyContinuousTimeComputations}}
  \end{document}
}

\title{A Survey on Continuous Time  Computations}

\author{Olivier Bournez\inst{1,2}
\and Manuel L. Campagnolo\inst{3,4}
\institute{ INRIA Lorraine,
\and LORIA (UMR 7503 CNRS-INPL-INRIA-Nancy2-UHP),
Campus scientifique, BP 239, 54506 Vand{\oe}uvre-L{è}s-Nancy, France,
\\ \email{Olivier.Bournez@loria.fr}
\and DM/ISA, Technical University of Lisbon, Tapada da Ajuda, 1349-017 Lisboa, 
Portugal, 
\and SQIG/IT  Lisboa, \\ \email{mlc@math.isa.utl.pt}
}}
\maketitle


\begin{abstract}
    We provide an overview of  theories of continuous
    time computation.  These theories 
    allow us to understand both the
    hardness of questions related to continuous time dynamical
    systems and the computational power of continuous
    time analog models.  
    We survey the
    existing models, summarizing results, and point to relevant references in the literature.  
\end{abstract}

\section{Introduction}

Continuous time  systems arise 
as soon as one attempts 
to model
systems that evolve over a continuous space with a continuous time.
They can even 
emerge 
 as natural descriptions of discrete time or space
systems. Utilizing continuous time systems is a common approach in fields such as biology, physics
or chemistry, when  a huge population of agents (molecules,
individuals, \dots) is abstracted into real quantities such as 
proportions or thermodynamic data \cite{HSD03}, \cite{Murray-VolI}.

There are several approaches that have led to theories on continuous time
computations. We will explore in greater depth two primary approaches.
One, which we call {\em inspired by continuous time analog machines}, has
its roots in models of natural or artificial analog machinery. The
other, we refer to as {\em inspired by continuous time system
theories}, is broader in scope. It comes from research on continuous time systems theory from a computational perspective. Hybrid systems and automata theory, for example, are two sources.

A wide range of problems related to theories of continuous time computations are encompassed by these two approaches. They
originate in fields as diverse as verification (see e.g.
\cite{AMP95}), control theory (see e.g. \cite{Bra95}), VLSI design (see
e.g. \cite{Mills95}, \cite{Mills05}), neural networks (see for example
\cite{Orponen}) and recursion theory on the reals
 (see e.g. \cite{Moo95b}).

At its beginning, continuous time computation theory was mainly 
concerned with analog machines. 
Determining which systems can actually be considered
as computational models is a very intriguing question. This relates to
the philosophical discussion about what is a programmable machine, which is 
beyond the scope of this chapter.
Nonetheless, 
there are some early examples of built analog devices that are generally accepted as 
programmable
machines.  They include Bush's landmark 1931 Differential Analyzer
\cite{Bush31}, 
as well as 
Bill Phillips' Finance Phalograph, 
Hermann's 1814 Planimeter, Pascal's 1642 Pascaline, or even the
87 b.c. Antikythera mechanism: see \cite{AnalogComputerMuseum}.
Continuous time computational models also include neural networks
and systems that can be built using electronic
analog devices. Since continuous time systems 
are conducive to modeling huge populations, one might speculate that they will have a
prominent role in 
analyzing massively parallel systems such as the
Internet \cite{Pap01}.

The first true model of a universal continuous time machine was proposed 
  by Shannon \cite{Sha41}, who introduced it as a model of the
  Differential Analyzer. 
  During the 1950s and 60s an extensive body of literature was published 
  about the programming of such
  machines\footnote{See for example the very instructive Doug Coward
    's web \textit{Analog Computer Museum} \cite{AnalogComputerMuseum}
    and its bibliography. 
    This literature reveals a quite
    forgotten art of programming continuous time and hybrid (digital-analog) machines, which level of sophistication is close to today's engineering programming.}. 
    There were also a number of significant publications on 
  how to use analog devices to solve
  discrete or continuous problems: see e.g. \cite{VSD86} and the
  references therein.
  However, most of this early  literature is 
  now only marginally relevant 
  given the ways in which our current
  understanding of computability and complexity theory have developed.


  The research on artificial neural networks, despite the fact that it mainly focused on discrete time analog models, has motivated a change of perspective
 due to its many shared  
  concepts and goals with today's standard computability and complexity theory \cite{Orponen},
  \cite{OrponenComplexity}. Another line of development of continuous time computation theory has been motivated by
 hybrid systems, particularly by questions related to the
  hardness of their verification and control: see for example 
  \cite{Bra95} and \cite{AMP95}. 
In recent years there has also been a surge of interest in 
alternatives to classical digital models other than continuous time systems.
Those alternatives include discrete-time analog-space 
models
like  artificial neural networks \cite{Orponen},
optical models \cite{Woods05}, signal  machines \cite{Durand-Lose05}
and the Blum Shub and Smale model \cite{BCSS}. More generally there have 
also been many recent developments in non-classical and more-or-less realistic or futuristic
models such as exotic cellular automata models \cite{GM04}, 
molecular or natural computations \cite{Head87}, \cite{Adl94},
\cite{Lipton94},
\cite{MembraneBook}, black hole computations
\cite{Hogarth92}, or 
  quantum computations 
 \cite{Deu85}, \cite{Gru97}, \cite{Sho94}, 
  \cite{Kieu04}. Some of these contributions are detailed in this volume. 


  The computational power of discrete time models are fairly well known and
  understood thanks in large part to the Church-Turing thesis. The Church-Turing thesis states that all reasonable
  and sufficiently powerful models are equivalent. For continuous time
  computation, the situation is far from being so clear, and there has not been a significant effort toward unifying concepts. Nonetheless,  some recent 
results establish the equivalence between apparently distinct models 
\cite{GC03}, \cite{Gra04}, \cite{dsg05}, and \cite{BCGH07}, which give us hope that a unified theory of continuous time computation may not be too far in the future.

  This text can be considered an up-to-date version of Orponen's
  1997 survey \cite{Orponen}. Orponen states at the end of his introduction of that the
  effects of imprecision and noise in analog computations are still far
  from being understood and  that a robust complexity theory of
  continuous time models has yet to be developed. Although this evaluation remains largely accurate with regard to imprecision and noise, we will see in the present survey that in the intervening decade much progress in understanding the computability and even the complexity of continuous time computations has been made.


  This chapter is organized as follows. In Section 
\ref{sec:deux}, we 
  review the most relevant continuous time models. In sections \ref{sec:trois} and \ref{sec:quatre} we discuss, respectively, computability and complexity issues in continuous time computations. In these sections we focus mainly on continuous time dynamical systems. In Section \ref{sec:cinq} we address the effect of imprecision and noise in analog computations. Finally, in Section~\ref{sec:six}, we conclude with some general insights and directions of further research in the field of  continuous time computation.

\section{ Continuous Time Models} 
\label{sec:deux}

With a historical perspective in mind, we outline in this section
several of 
the major classes of continuous time models 
that motivated interest
in this field
These models also illustrate concepts like continuous dynamics and input/output.


\subsection{Models inspired by analog machines}

\subsubsection{GPAC and other circuit models}

Probably, the  best known universal continuous time machine
is the \textit{Differential Analyzer}, built at MIT under the
supervision of Vannevar Bush \cite{Bush31} for the first time in
1931. The idea of assembling integrator devices to solve differential
equations dates back to Lord Kelvin in 1876 \cite{Tho76}.
Mechanical\footnote{And even \textit{MECANO} machines: see
  \cite{Bow96}.}, and later on electronic, differential
analyzers were used to solve 
various kinds of 
differential
equations primarily related to problems in the field of engineering:
see for e.g. \cite{Bow96}, or more generally \cite{IEEEAnnals} for
historical accounts.  By the 1960s differential analysers were
progressively discarded in favor of digital technology.

\begin{figure}[p]
\begin{center}%
\setlength{\unitlength}{1300sp}%
\begingroup\makeatletter\ifx\SetFigFont\undefined%
\gdef\SetFigFont#1#2#3#4#5{%
  \reset@font\fontsize{#1}{#2pt}%
  \fontfamily{#3}\fontseries{#4}\fontshape{#5}%
  \selectfont}%
\fi\endgroup%
\begin{picture}(9038,4512)(1501,-5461)
{\put(4200,-1561){\line( 1, 0){600}}}%
{\color[rgb]{0,0,0}\put(10500,-2161){\framebox(1800,1200){}}}%
{\color[rgb]{0,0,0}\put(2400,-4861){\framebox(1800,1200){}}}%
{\color[rgb]{0,0,0}\put(10500,-4861){\framebox(1800,1200){}}}%
{\color[rgb]{0,0,0}\put(9900,-1261){\line( 1, 0){600}}}%
{\color[rgb]{0,0,0}\put(9900,-1861){\line( 1, 0){600}}}%
{\color[rgb]{0,0,0}\put(12300,-1561){\line( 1, 0){600}}}%
{\color[rgb]{0,0,0}\put(12300,-4261){\line( 1, 0){600}}}%
{\color[rgb]{0,0,0}\put(4200,-4261){\line( 1, 0){600}}}%
{\color[rgb]{0,0,0}\put(1800,-3961){\line( 1, 0){600}}}%
{\color[rgb]{0,0,0}\put(1800,-4561){\line( 1, 0){600}}}%
{\color[rgb]{0,0,0}\put(9900,-3961){\line( 1, 0){600}}}%
{\color[rgb]{0,0,0}\put(9900,-4561){\line( 1, 0){600}}}%
\put(3151,-1636){\makebox(0,0)[lb]{\smash{$k$}}}
\put(4951,-1636){\makebox(0,0)[lb]{\smash{$k$}}}
\put(9601,-1261){\makebox(0,0)[lb]{\smash{$u$}}}
\put(9601,-1936){\makebox(0,0)[lb]{\smash{$v$}}}
\put(11300,-1636){\makebox(0,0)[lb]{\smash{$+$}}}
\put(13000,-1636){\makebox(0,0)[lb]{\smash{$u+v$}}}
\put(1500,-3961){\makebox(0,0)[lb]{\smash{$u$}}}
\put(1500,-4636){\makebox(0,0)[lb]{\smash{$v$}}}
\put(3151,-4261){\makebox(0,0)[lb]{\smash{$\int$}}}
\put(4951,-4336){\makebox(0,0)[lb]{\smash{$w \left\{ \begin{array}{lll} 
w'(t) & =& u(t)v'(t) \\ w(t_0) & = & \alpha \\ \end{array} \right.$}}}
\put(13000,-4336){\makebox(0,0)[lb]{\smash{$uv$}}}
\put(11300,-4261){\makebox(0,0)[lb]{\smash{$×$}}}
\put(9601,-3961){\makebox(0,0)[lb]{\smash{$u$}}}
\put(9601,-4636){\makebox(0,0)[lb]{\smash{$v$}}}
\put(2401,-2800){\makebox(0,0)[lb]{\smash{A constant unit}}}
\put(10501,-2800){\makebox(0,0)[lb]{\smash{An adder unit}}}
\put(2401,-5500){\makebox(0,0)[lb]{\smash{An integrator unit}}}
\put(10501,-5500){\makebox(0,0)[lb]{\smash{A multiplier unit}}}
{\color[rgb]{0,0,0}\put(2401,-2161){\framebox(1800,1200)}}
\end{picture}%
\end{center}
\caption{Different types of units used in a GPAC.}
\label{fig:units}
\end{figure}
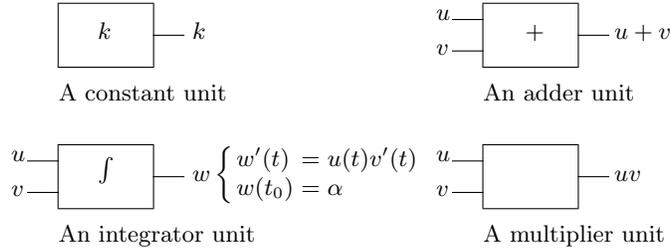

The first theoretical study of the computational capabilities of
continuous time universal machines was published by Shannon. In
\cite{Sha41}, he 
 proposed
 what is now referred to as the \textit{General Purpose Analog
  Computer (GPAC)} as a theoretical model of Vannevar Bush's
differential analyzer. The model, later refined in the series of 
papers
\cite{PER74}, \cite{LR87}, \cite{GC03}, \cite{Gra04}, consists 
of families of
circuits built with the basic units presented in Figure
\ref{fig:units}. 
There are some restrictions to the kinds of interconnectivity that are allowed to avoid undesirable behavior:
\textsl{e.g.}
non-unique outputs. For further details and discussions, refer to
\cite{GC03} and \cite{Graca}.

\begin{figure}[p]
\begin{center}
\begin{picture}(350,60)(13,10)
\put(40,20){\framebox(30,20){$\int$}}
\put(100,16){\framebox(30,20){$\int$}} \put(100,30){\line(-1,0){30}}
\put(160,20){\framebox(30,20){$\int$}} \put(160,26){\line(-1,0){30}}
\put(105,45){\framebox(20,15){\textrm{-1}}}
\put(160,34){\line(-1,0){15}} \put(145,34){\line(0,1){18.5}}
\put(145,52.5){\line(-1,0){20}} \put(40,26){\line(-1,0){20}}
\put(30,26){\circle*{2}} \put(30,26){\line(0,-1){16}}
\put(100,22){\line(-1,0){10}} \put(90,22){\line(0,-1){12}}
\put(90,10){\line(-1,0){60}} \put(210,30){\line(-1,0){20}}
\put(200,30){\circle*{2}} \put(200,30){\line(0,1){40}}
\put(200,70){\line(-1,0){170}} \put(30,70){\line(0,-1){36}}
\put(40,34){\line(-1,0){10}} \put(16,24){$\scriptstyle t$}
\put(212,29){$\scriptstyle y_{3}$} \put(140,21){$\scriptstyle
y_{2}$} \put(80,33){$\scriptstyle y_{1}$} \put(240,40){$\left\{
\begin{array}
[c]{lll}
y_{1}^{\prime}=y_{3} & \& & y_{1}(0)=1\\
y_{2}^{\prime}=y_{1} & \& & y_{2}(0)=0\\
y_{3}^{\prime}=-y_{1} & \& & y_{3}(0)=0
\end{array}
\right.  $}
\end{picture}
\end{center}
\caption{Generating $\cos$ and $\sin$ via a GPAC: circuit version on
the left and ODE version on the right. One has $y_1=\cos$,
$y_2=\sin$, $y_3=-\sin$.}
\label{fig:sincos}
\end{figure}
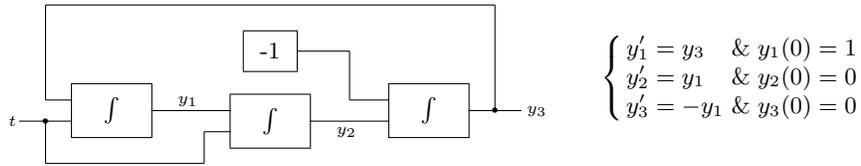

Shannon, in his original paper, already mentions that the GPAC
generates polynomials, the exponential function, the usual
trigonometric functions, and their inverses (see Figure~\ref{fig:sincos}). More generally, he 
claimed in \cite{Sha41} that a function can be generated by a GPAC iff
it is differentially algebraic. i.e. it satisfies some algebraic
differential equation of the form $$p\left(
  t,y,y^{\prime},...,y^{(n)}\right) =0,$$ where $p$ is a non-zero
polynomial in all its variables.
As a corollary, and noting that the Gamma function
$\Gamma(x)=\int_{0}%
^{\infty}t^{x-1}e^{-t}dt$ or the Riemann's Zeta function
$\zeta(x)=\sum_{k=0}^\infty \frac1{k^x}$ are not d.a. \cite{Rub89a}, it
follows that the Gamma and the Zeta functions are examples of
functions that can not be generated by a GPAC.

However, Shannon's proof relating functions generated by GPACs with
differentially algebraic functions was incomplete (as pointed out and
partially corrected by \cite{PER74}, \cite{LR87}). However, for the
more robust class of GPACs defined in \cite{GC03}, the following
stronger property holds: a scalar function $f: \mathbb{R} \to\mathbb{R}$ is 
generated
by a GPAC iff it is a component of the solution of a system $y^{\prime}=p(t,y)$,
where $p$ is a vector of polynomials. A function $f: \mathbb{R} \to
\mathbb{R}^{k}$ is generated by a GPAC iff all of its components are.

 The $\Gamma$ function is
  indeed GPAC computable, if a notion of computation inspired from
  recursive analysis is considered \cite{Gra04}. GPAC computable
  functions in this sense correspond precisely to computable functions
  over the reals \cite{BCGH07}.  

Rubel proposed \cite{Rub93} an extension of 
Shannon's original GPAC. In Rubel's model, 
the Extended Analog Computer (EAC),   operations to solve boundary value problems or
to take certain infinite limits were added.
%
We refer
to \cite{Mills95} and \cite{Mills05} for descriptions of  
actual
working implementations of Rubel's EAC.

More broadly, a discussion of circuits made of general basic
units has been presented recently in \cite{TZ07}. 
Equational specifications of such circuits, as well as their semantics, are given by fixed points of operators over the space of continuous streams.  
Under suitable hypotheses, this operator is contracting and
an extension of Banach fixed point theorem for metric spaces
guarantees existence and unicity of the fixed point. Moreover, that
fixed point can also be proved to be continuous and
\textit{concretely} computable whenever the basic modules 
also have those properties.

\subsubsection{Hopfield network models}
\label{sec:liap}

Another well known continuous time model is the 
``neural network''
model proposed by John Hopfield in 1984 in \cite{Hopfield84}. 
These 
networks
can be implemented in electrical \cite{Hopfield84} or optical  
hardware \cite{Stoll88}.

A 
symmetric 
Hopfield network is made of a finite number, say $n$, of
simple computational units, or \textit{neurons}. The architecture of
the network is given by some (non oriented) graph whose nodes are the
neurons, and whose edges are labeled by some weights, the
\textit{synaptic weights}. The graph can be assumed to be complete by
replacing the absence of a connection between two nodes by an edge
whose weight is null.

The state of each neuron $i$ at time $t$ is given by some real value
$u_i(t)$. Starting from some given initial state $\vu_0 \in \R^n$, the
global dynamic of the network is defined by a system of differential
equations
$$ C_i u_i'(t) = \sum_j W_{i,j} V_j - u_i/R_i + I_i,$$ where
$V_i=\sigma(u_i)$, $\sigma$ is some saturating function such as 
$\sigma(u)=\alpha \tan u
+ \beta$, $W_{i,j}=W_{j,i}$ is the weight of the edge between $i$ and
$j$, $C_i,I_i,R_i$ are some constants \cite{Hopfield84}.

Hopfield proved in \cite{Hopfield84}, by a Lyapunov-function argument,
that such systems are \textit{globally asymptotically stable},
i.e. from any initial state, the system relaxes toward some stable
equilibrium state.  Indeed, consider for example the energy function
\cite{Hopfield84}
$$E = -\frac12 \sum_{i} \sum_j W_{i,j} V_i V_j + \sum_i \frac1{R_i} 
\int_0^{V_i} \sigma^{-1}(V) dV + \sum_i I_i V_i.$$ The function $E$ is bounded and its 
derivative is negative. Hence the time evolution 
of the 
whole system is a motion in a state space that seeks out (possibly 
local) minima of $E$.

This convergence behavior has been used by Hopfield to explore various
applications such as associative memory, or to solve combinatorial
optimization problems \cite{Hopfield84},
\cite{Ref12-VSD86}.



An exponential lower bound on the convergence time of 
continuous time Hopfield networks has been related to 
their dimension in \cite{SimOrp03b}. 
Such 
continuous time symmetric networks can be proved to simulate any finite
binary-state discrete-time recurrent neural network \cite{SO2000},
\cite{SO2003}.


\subsubsection{Networks of spiking neurons} 
If one classifies, following \cite{Maass:97b},  neural network models   
according to 
their activation functions and dynamics,
three different generations can be distinguished. The first
generation, with discontinuous activation functions, includes
multilayer perceptrons, 
Hopfield networks,
 and
Boltzmann machines (see for example \cite{AbdiNN} for an introduction  
to all mentioned Neural Networks models).
The output of this generation of networks is digital. The second  
generation of networks use continuous activation functions instead of
step or threshold functions to compute the output signals.
These include feedforward and recurrent
sigmoidal neural network,  radial basis functions networks, and  
continuous time Hopfield networks. Their
input and output is analog. The third generation of networks is based on
spiking neurons and encodes variables in time differences between
pulses.  This generation
exhibits continuous time
dynamics and is the most biologically realistic \cite{MB98}.

There are several mathematical models of spiking neurons of which we will focus on one,
whose computational properties
have been 
investigated in depth. The Spiking Neural Network model 
is represented by a finite directed graph. To each node $v$ (neuron) of the
graph is associated a \textit{threshold function} $\theta_v: \RP \to \R \cup
\{\infty\}$, and to each edge $(u,v)$ (synapse) is associated a
\textit{response-function} $\epsilon_{u,v} :\RP \to \R$ and a \textit{weight-function}
$w_{u,v}$.

  For a non-input neuron $v$, one defines its set $F_v$ of {\em firing
  times} recursively. The first element of $F_v$ is $\inf\{t | P_v(t)
    \geq \theta_v(0)\}$, and for any $s \in F_v$, the next larger
  element of $F_v$ is $\inf\{t | t>s \mbox{ and } P_v(t)
    \geq \theta_v(t-s)\}$ where
$$P_v(t) = 0+ \sum_{u} \sum _{s \in F_u, s <t} w_{u,v}(s) 
\epsilon_{u,v}(t-s).$$

The $0$ above can be replaced by some bias function.  We use it here
to guarantee that $P_v$ is well-defined even if $F_u=\emptyset$ for
all $u$ with $w_{u,v} \neq 0$.  To approximate biological realism,
restrictions are placed on the allowed response-functions and
bias-functions of these models: see \cite{Maass:97b},
\cite{Maass:98f}, \cite{NM2001}, or \cite{Maass:02}, \cite{Maass:03},
for discussions on the model.  In particular, rapidly fading memory is
a biological constraint which prevents chaotic behavior in networks
with a continuous time dynamic.  Recently, the use  of feedback to
overcome the limitations of such a constraint was analized in
\cite{MJS07}.
 
The study of the computational power of several variants of spiking neural networks was
initiated in 
\cite{Maa96b}.
Noisy extensions of the model have been considered 
\cite{Maa96c}, \cite{Maass:97c}, 
\cite{MN2000}. A survey of complexity results can be found in 
\cite{SOSurvey}.  Restrictions that are easier to implement in hardware 
versions
have also been investigated in 
\cite{MR97b}.  



\subsubsection{$\R$-recursive functions } \label{sec:syntax} 

\newcommand{\rrec}{\mathcal{M}}
\newcommand{\rprim}{\mathcal{I}}
\newcommand{\relem}{\mathcal{L}}
\newcommand{\comp}{\mathsf{comp}}
\newcommand{\minim}{\mathsf{minim}}
\newcommand{\integ}{\mathsf{int}}
\newcommand{\li}{\mathsf{LI}}
\newcommand{\lims}{\mathsf{lim}}

Moore proposed a theory of recursive functions on the reals in
\cite{Moo95b}, which is defined in analogy with classical recursion
theory and corresponds  to a conceptual analog computer operating
in continuous time.  As we will see, this continuous time model has in
particular the capability of solving differential equations, similar
to an idealized analog integrator of the GPAC.  In fact, the theory of
$\R$-recursive functions can be seen as an extension of Shannon's
theory for the GPAC. A general discussion of the motivations behind
$\R$-recursion theory can be found in \cite{MC05Eatcs}.

A function algebra $[B_1,B_2,...;O_1,O_2,...]$ is the smallest set
containing basic functions $\{B_1,B_2,...\}$ and closed under certain
operations $\{O_1,O_2,...\}$, which take one or more functions in the
class and create new ones.  Although function algebras have been
defined in the context of recursion theory on the integers, and widely
used to characterize computability and complexity classes
\cite{Clo95}, they are equally suitable to define classes of real
valued recursive functions.

The $\R$-recursive functions were first defined in
\cite{Moo95b}. These are given by the function algebra $\rrec=[0,1,U;
\comp, \integ, \minim]$,\footnote{We consider that the operator
  $\integ$ preserves analyticity (see \cite{mlc:moo:fgc:99a},
  \cite{cam:umc:02}).} where $U$ is the set of projection functions
$U_i(\vec{x})=x_i$, $\comp$ is composition, $\integ$ is an operation
that given $f$ and $g$ returns the solution of the initial value
problem $h(\vec{x},0)=f(\vec{x})$ and $\partial_y
h(\vec{x},y)=g(\vec{x},y,h)$ and $\minim$ returns the smallest zero
$\mu_y f(\vec{x},y)$ of a given $f$.  Moore also studied the weaker
algebra $\rprim=[0,1,-1,U; \comp, \integ]$ and claimed its
equivalence with the class of unary functions generated by the GPAC
\cite{Moo95b}.

Many non recursively enumerable sets are $\R$-recursive. Since
$\minim$ is the operation in $\rrec$ which gives rise to uncomputable
functions, a natural question is to ask if $\minim$ can be replaced by
some other operation of mathematical analysis. This was done in
\cite{MC04a} where $\minim$ is replaced by the operation $\lims$,   which returns the infinite limits of the functions in the algebra. These authors stratify $[0,1,-1,U; \comp,
\integ, \lims]$ according to the allowed number ($\eta$) of nested
limits and relate the resulting $\eta$-hierarchy with the arithmetical
and analytical hierarchies. In \cite{LCM07} it is shown that the
$\eta$-hierarchy does not collapse (see also \cite{Loff07}), which
implies that  infinite limits and first order integration are
not interchangeable operations \cite{LCM07b}.

The algebra $\rprim$ only contains analytic functions and is not
closed under iteration \cite{mlc:moo:fgc:99a}. However, if an
arbitrarily smooth extension to the reals $\theta$ of the Heaviside
function is included in the set of basic functions of $\rprim$, then
$\rprim+\theta$ contains extensions to the reals of all primitive
recursive functions.  

The  closure of fragments of $\rprim+\theta=[0,1,-1,\theta, U;
\comp, \integ]$ under discrete operations like 
bounded
products, bounded sums and bounded recursion, has been investigated in
the thesis \cite{Campagnolo} and also in the papers \cite{cam:moo:fgc:00}, \cite{cam:umc:02}, \cite{cam:tcs:04}.

In particular, several
authors studied the function algebra $\relem=[0,1,-1,\pi,\theta, U;$
$\comp, \li]$ where the $\li$ is only able to solve {\em linear}
differential equations (i.e., it restricts $\integ$ to the case
$\partial_y h(\vec{x},y)=g(\vec{x},y)\, h(\vec{x},y)$). The class
$\relem$ contains extensions to the reals of all the elementary
functions, 
\cite{cam:moo:fgc:00}.

Instead of asking which computable functions over $\N$ have extensions
to $\R$ in a given function algebra, Bournez and Hainry consider
classes of functions over $\R$ computable according to  recursive
analysis, and characterize them precisely with function algebras. 
This was done for the elementarily computable functions
\cite{BH05}, characterized as $\relem$ closed under a restricted limit
schema. This was extended to yield a characterization of the whole
class of computable functions over the reals
\cite{FundamentaInformatica2006}, adding a restricted minimisation schema.
Those results provide syntactical characterizations of real computable
functions in a continuous setting, which is arguably more natural than
the higher order Turing machines of recursive analysis.  

A more general approach to the structural complexity of real recursive
classes, developed in \cite{cam:oja:06}, is based on the notion of
approximation.  This notion was used to lift complexity results from $\N$ to $\R$, and was applied in particular to characterize $\relem$.

Somewhat surprisingly, the results above indicate that two distinct models of computation over the reals (computable analysis and real recursive functions) can be linked  in an elegant way. 

\subsection{Models inspired by continuous time system theories} 

\subsubsection{Hybrid Systems} 
An increasing number of systems exhibit some interplay between
discrete and analog behaviors. The investigation of these systems has 
led to relevant new  results about continuous time computation.

A   variety of models have been considered: see for example the
conference series \textit{Hybrid Systems Computation and Control} or \cite{Bra95Thesis}. However,  hybrid
systems\footnote{``Hybrid'' refers here to the fact that the systems
  have intermixed discrete and continuous evolutions. This differs 
  from  historical literature about analog computations,
  where ``hybrid'' often refers to machines with a mixture of analog and digital
  components.} are essentially
modeled either as differential equations with
discontinuous right hand sides, as differential equations with
continuous and discrete variables, or as hybrid automata. A hybrid
automaton is a finite state automaton extended with variables. Its 
associated dynamics consists of guarded discrete transitions between states 
of the automaton that can reset some variables. Typical
properties of hybrid systems that have been considered are reachability, 
stability and
controllability.


With respect to the differential equation modeling approach, 
Branicky
proved in \cite{Bra95} that any hybrid system model that can
implement a clock and also implement general continuous ordinary differential
equations is able to simulate Turing machines.  Asarin, Maler and
Pnueli proved in \cite{AMP95} that piecewise constant differential
equations can simulate Turing machines in $\R^3$, while the
reachability problem for these systems in dimension $d \leq 2$ is
decidable \cite{AMP95}.  Piecewise constant differential equations, as
well as many hybrid systems models, exhibit the so-called \textit{Zeno's
phenomenon}: an infinite number of discrete transitions may happen in
a finite time. This has been used in \cite{AM95} to prove that
arithmetical sets can be recognized in finite time by these
systems. Their exact computational power has been characterized in
terms of their dimension in \cite{Bou97b} and \cite{These}. The
Jordan's theorem based argument of \cite{AMP95} to get decidability
for planar piecewise constant differential equations has been
generalized for planar polynomial systems
 \cite{CV95} and for planar differential inclusion systems 
\cite{Asarin:2001:DRP}.

There is extensive literature on the hybrid automata modeling approach
about determining the exact frontier between decidability and
non-decidability for reachability properties, according to the type of
allowed dynamics, guards, and resets.  The reachability property has
been proved decidable for timed automata \cite{AD91}.
By reduction to this result, or by a finite bisimulation argument in the same spirit, this has
also been generalized to multirate automata \cite{ACH+95}, to specific
classes of updatable timed automata in \cite{BouyerEtAl00},
\cite{BouyerEtAl00a}, and to initialized rectangular automata in \cite{HKPV95},
\cite{PV94}. 
There is a multitude of undecidability results,
most of which rely on simulations of Minsky two-counter machines.  For
example, the reachability problem is semi-decidable but non-decidable
for linear hybrid automata \cite{ACH+95}, \cite{NOSY93}.  The same
problem is known to be undecidable for rectangular automata with at
least $5$ clocks and one two-slope variable \cite{HKPV95}, or for
timed automata with two skewed clocks \cite{ACH+95}. For discussion of these results, see also \cite{AsaSch02}.   Refer to \cite{BT99a}, and \cite{Collins04} or to the survey 
\cite{BTSurvey_} for properties other than reachability 
(for example, stability and observability).

\textit{O-minimal} hybrid systems are initialized hybrid systems whose
relevant sets and flows are definable in an o-minimal theory. 
These systems 
always admit a finite bisimulation \cite{GLP00}.  However, their
definition can be extended to a more general class of
``non-deterministic'' o-minimal systems \cite{BM05}, for which the
reachability problem is undecidable in the Turing model, as well as in
the Blum Shub Smale model of computation \cite{Bri06}.  Upper bounds
have been obtained on the size of the finite bisimulation for
Pfaffian hybrid systems \cite{KorovinaV04} \cite{KorovinaV06} using
the word encoding technique introduced in \cite{BM05}.

\subsubsection{Automata theory} There have been several attempts to
adapt classical discrete automata theory to continuous time; this is
sometimes referred to 
as the general program of Trakhtenbrot
\cite{Trakhtenbrot95}.

One attempt is related to 
timed automata, which can be seen as languages recognizers \cite{AD94}. 
Many specific decision problems have
been considered for timed automata: see survey
\cite{AlurM04}. 
Timed regular languages are known to be closed under intersection, union,
and renaming, but not under complementation. The membership and empty
language problems are decidable, whereas inclusion and universal
language problems are undecidable.  The closure of timed regular languages under shuffling is
investigated in \cite{Finkel06}.  Several variants of Kleene's theorem
are  established  \cite{ACM96}, \cite{Asa_},
\cite{AsaCasMal02}, \cite{BouPet99}, \cite{BouyerP02}, and in
\cite{AsarinD02}. There have been some attempts to establish
pumping lemmas \cite{Beauquier98}. A review, with discussions and open
problems related to this approach can be found in \cite{Asarin04}.
 
An alternative and independent automata theory over continuous time has been 
developed in \cite{RabTra97},
\cite{Trakhtenbrot99}, \cite{Rabinovich03b}. Here automata are not
considered as language recognizers, but as computing operators on
signals. A signal is a function from the non-negative real numbers to
a finite alphabet (the set of the channel's states). Automata theory
is extended to continuous time, and it is argued that the behavior of
finite state devices is ruled by so-called finite memory retrospective
functions. These are proved to be
speed-independant, i.e. independant under ``stretchings'' of the time
axis. Closure properties of operators on signals are established, and
the representation of finite memory retrospective functions by finite
transition diagrams (transducers) is discussed. See also
\cite{Franc02} for a detailed presentation of Trakhtenbrot and
Rabinovich's theory, and for discussions about the representation of
finite memory retrospective operators by circuits.

Finally, another independent approach is considered in \cite{Ruo04}, 
where 
Chomsky-like hierarchies are established for families of sets of 
piecewise
continuous functions. Differential equations, associated to specific 
memory structures, are used to recognize sets of functions. 
Ruohonen shows that the resulting hierarchies are not trivial and 
establishes closure properties and
inclusions between classes.


\subsection{Other computational models}

In addition to the two previously described approaches, there are a
number of other computational models that have lead to interesting
developments in continuous time computation theory.

The question of whether Einstein's general relativity equations admit
space-time solutions that allow an observer to view an eternity in a
finite time was investigated and proved possible  in
\cite{Hogarth92}. The question of whether this implies that
super-tasks can in principle be solved has been investigated in 
\cite{EN93}, \cite{HogarthThesis}, \cite{HogarthPSA94},
\cite{HogarthTrouNoir06}, \cite{TrouNoir}, \cite{NA06}, \cite{ND06},
and \cite{Welch06}.

Some machine inspired models 
are neither clearly digital nor analog. 
For example, the power of planar mechanisms
attracted great interest in England and France in
the late 1800s, and in the 1940s in Russia. 
Specifically, these consisted of rigid bars constrained to a plane and joined at either end by
rotable rivets. 
A theorem
attributed\footnote{The theorem is very often attributed to Kempe
  \cite{Ref3-Smi98}, \cite{Ref88-Smi98}, even if he apparently never
  proved exactly that.}  to Kempe \cite{Kempe76} states that they are
able to compute all algebraic functions: see for
  e.g. \cite{Ref3-Smi98} or \cite{Ref88-Smi98}.

\section{ODEs and properties}
\label{sec:trois}

Most of the continuous time models described above have a continuous
dynamics described by differential equations.  In Shannon's GPAC and
Hopfield networks, the input corresponds to the initial condition
while the output is, respectively, the time evolution or the
equilibrium state of the system. Other models are language
recognizers. The input again corresponds to the initial condition, or some
initial control, and the output is determined by some accepting region
in the state space of the system. All these systems therefore fall 
into the framework of dynamical systems. 

In this section we will recall some fundamental results about
dynamical systems and differential equations and discuss how
different models can be compared in this general framework.

\subsection{ODEs and dynamical systems}

Let's consider that we are working in $\R^n$ (in general, we could
consider any vector space with a norm). Let us consider $f: E \to \R^n$, where $E
\subset \R^n$ is open. An ODE is given by $y^\prime=f(y)$ and its
solution is a differentiable function $y: I \subset \R \to E$ that
satisfies the equation.

For any $x \in E$, the fundamental existence-uniqueness theorem (see
  e.g. \cite{HSD03}) for differential equations states that if
$f$ is Lipschitz on $E$, i.e. if there exists $K$ such that
$||f(y_1)-f(y_2)||<k||y_1-y_2||$ for all $y_1,y_2 \in E$, then the
solution of
\begin{equation}
y^\prime=f(y), \qquad y(t_0)=x
\label{eq:ivp}
\end{equation}
exists and is unique on a certain maximal interval of existence $I
\subset \R$.  In the terminology of dynamical systems, $y(t)$ is
referred to as the \textit{trajectory}, $\R^n$ as the \textit{phase
  space}, and the function $\phi(t,x)$, which gives the position
$y(t)$ of the solution at time $t$ with initial condition $x$, as the
\textit{flow}. The graph of $y$ in $\R^n$ is called the
\textit{orbit}.

  In particular, if $f$ is continuously differentiable on $E$ then the
  existence-uniqueness condition is fulfilled \cite{HSD03}. 
  Most of the mathematical theory has been developed in this case, but
  can be extended to weaker conditions. In particular, if $f$ is
  assumed to be only continuous, then uniqueness is lost, but
  existence is guaranteed: see for example \cite{CL72}. If $f$ is
  allowed to be discontinuous, then the definition of solution needs to be
  refined. This is explored by Filippov in 
  \cite{FilippovBook}. 
  Some hybrid system models use distinct and ad hoc notions of
  solutions. For example, a solution of a piecewise constant
  differential equation in \cite{AMP95} is a continuous function whose
  right derivative satisfies the equation.


  In general, a dynamical system can be defined as the action of a
  subgroup $\mathcal{T}$ of $\R$ on a space $X$, i.e. by a function (a
  flow) $\phi: \mathcal{T} \times X \to X$ satisfying the following
  two equations
\begin{equation}
\phi(0,x)=x \label{flow:one}
\end{equation}
\begin{equation}
\phi(t,\phi(s,x))=\phi(t+s,x). \label{flow:two}
\end{equation}

It is well known that subgroups $\mathcal{T}$ of $\R$ are either dense
in $\R$ or isomorphic to the integers. In the first case, the time is
called continuous, in the latter case, discrete.

Since flows obtained by initial value problems (IVP) of the form
(\ref{eq:ivp}) satisfy equations (\ref{flow:one}) and (\ref{flow:two}), they correspond to specific
continuous time and space dynamical systems. Although not all continuous time
and space dynamical systems can be put in a form of a differential
equation, IVPs of the form (\ref{eq:ivp}) are sufficiently general
to cover a very wide class of such systems. In particular, if $\phi$ is 
continuously differentiable, then $y^\prime=f(y)$, with $f(y) = \left. 
\frac{d}{dt} \phi(t,y) \right|_{t=0}$, describes the dynamical system. 

For discrete time systems, we can assume without loss of generality
that $\mathcal{T}$ is the integers.  The analog of of IVP (\ref{eq:ivp}) 
for discrete time
systems is a recurrence equation of type 
\begin{equation}
  y_{t+1}=f(y_t), \qquad y_0=x. \label{eq:ivpd}
\end{equation}

A dynamical system whose space is discrete and that evolves discretely
is termed digital, otherwise it is analog. A classification of some 
computational
models according to the nature of their space and time can be found in
Figure~\ref{fig:discont}.

\begin{center}
\begin{figure}[p] 
\hspace{-1cm}
\begin{tabular}{|ll||c|c|}
\hline
 & Space & Discrete & Continuous \\
Time &   &          &            \\
\hline
\hline
Discrete 
&&  \cite{Tur36}   machines     & Discrete time \cite{Hopfield84}  
neural networks    \\
&&  \cite{Church36} lambda calculus        & \cite{SS94} neural 
networks\\
&&  \cite{Kleene36} recursive functions                  & \cite{AMP95} 
PCD systems \\
&&   \cite{Post46} systems                & \cite{BSS89} machines \\  
&&    Cellular automata                & \cite{Woods05} optical machines 
\\
&&    Stack automata                & \cite{Durand-Lose05} signal 
machines \\
          &&  Finite State Automata             &  \cite{Moo96} 
Dynamical Recognizers \\
&&     \vdots                     &  \vdots \\
\hline
Continuous &&
\cite{BDEI} BDE models                          & \cite{Sha41} GPACs \\
            &&                & Continuous time \cite{Hopfield84} neural 
networks \\
           &&               & \cite{Bra95} hybrid systems \\
           &&               & \cite{AMP95} PCD systems \\
           &&               & \cite{AD91} timed automata \\
          &&               & \cite{Moo95b} $\R$-recursive functions \\
           &&               & \vdots  \\
\hline
\end{tabular}
\caption{A classification of some computational models, according to
  their space and time.}
  \label{fig:discont}
\end{figure}
\end{center}

\subsection{Dissipative and non-dissipative systems}

A point $x^*$ of the state space is called an \textit{equilibrium
  point} if $f(x^*)=0$. If the system is at $x^*$ it will remain
there. It is said to be \textit{stable} if for every neighborhood $U$
of $x^*$ there is a neighborhood $W$ of $x^*$ in $U$ such that every
solution starting from a point $x$ of $W$ is defined and is in $U$ for
all time $t>0$. The point is \textit{asymptotically stable} if, in
addition to the properties above, we have $\lim y(t)=x^*$
\cite{HSD03}.

Some local conditions on the differential $Df(x^*)$ of $f$ in $x^*$
have been clearly established. 
If at an equilibrium point $x^*$ all eigenvalues of
$Df(x^*)$ have negative real parts, then $x^*$ is asymptotically stable,
and furthermore nearby solutions approach $x^*$ exponentially.
In that case, $x^*$ is called a 
\textit{sink}. 
At
a stable equilibrium point $x^*$, no eigenvalue of $Df(x^*)$ can have
a positive real part \cite{HSD03}.

In practice, Lyapunov's stability theorem applies more broadly (i.e.,
even if $x^*$ is not a sink).  It states that if there exists a
continuous function $V$ defined on a neighborhood of $x^*$,
differentiable (except perhaps on $x^*$) with $V(x^*)=0$, $V(x)>0$ for
$x\neq x^*$, and $d V(x)/dt \leq 0$ for $x\neq x^*$ then $x^*$ is
stable. If, in addition, $dV(x)/dt < 0$ for $x\neq x^*$, then
$x^*$ is asymptotically stable: see \cite{HSD03}.

If the function $V$ satisfies the previous conditions everywhere, then
the system is \textit{globally asymptotically stable}. Whatever the
initial point $x$ is, the trajectories will eventually converge to
local minima of $V$.  In this context, the Lyapunov function $V$ can
be interpreted as an energy, and its minima correspond to attractors
of the dynamical system.  These are bounded subsets of the phase space
to which regions of initial conditions of nonzero volume converge as
time increases.

A dynamical system is called \textit{dissipative} if the volume of a set 
decreases under the flow for some region of the phase
space. Dissipative systems are characterized 
by the presence of attractors. 
By opposition, a dynamical
system is said to be \textit{volume-preserving} if the volume is conserved. For
instance, all Hamiltonian systems are volume-preserving because of
Liouville's theorem \cite{ArnoldMeca}. Volume preserving dynamical
system cannot be globally asymptotically stable \cite{ArnoldMeca}.

\subsection{Computability of solutions of ODEs}
\label{section:computabilityODEs}

Here we review some results on the computability of solutions of IVPs 
in the framework of recursive analysis (see e.g. \cite{LivreComputableAnalysis} and the corresponding chapter in this volume). 

In general, given a computable function $f$, one can investigate if
the solution of the IVP~(\ref{eq:ivp}) is also computable in the sense
of recursive analysis.  If we require that the IVP has a unique
solution then that solution is computable. Formally, if $f$ is computable on $[0,1] \times [-1,1]$ and
  the IVP $y^\prime=f(t,y)$, $y(0)=0$ has a unique solution on $[0,b]$, $0 < b
  \leq 1$, then the solution $y$ is computable on $[0,b]$.

This result also holds for a general $n$-dimensional IVP if its solution is unique
\cite{Ruo96}.
However, computability of solutions is lost as soon as uniqueness of
solutions is relaxed, even in dimension $1$. Indeed, the famous result of \cite{PouRic79}
shows that there exists a polynomial-time computable function $f:[0,1]
 \times [-1,1] \to \R$, such that the equation $y^\prime=f(t,y)$, with
$y(0)=0$,  has non-unique solutions, but none of them is
computable on any closed finite time interval.


Similar phenomena hold for other natural equations: the
$3$-dimensional wave equation (which is a partial equation), with
computable initial data, can have a unique solution which is nowhere
computable\footnote{However, in all these cases, the problems under
  study are ill-posed: either the solution is not unique, or it is
  unstable \ and the addition of some natural regularity conditions to
  prevent ill-posedness do yield computability \cite{WZ02}.}
\cite{PR81}, \cite{PZ97}.  Notice that, even if $f$ is assumed
computable and analytic, and the solution unique, it may happen that
the maximal interval $(\alpha,\beta)$ of existence of the solution is
non-computable \cite{dsg06a}. This same question is open if $f$ is polynomial. Those authors show, however, that if $f$
and $f^\prime$ are continuous and computable, then the solution of
$y^\prime=f(y,t), y(0)=x$, for computable $x$, is also computable on
its maximal interval of existence.
Refer also to \cite{PourEl:comap}, \cite{Ko83} for more
uncomputability results, and also to \cite{Ko83}, \cite{Ko91} for
related complexity issues.

\subsection{Static undecidability}

As observed in \cite{Asa95} and in \cite{Ruo97}, it is relatively simple but not very informative 
to get undecidability results with continuous time
dynamical systems, if $f$ encodes a undecidable problem. To illustrate 
this, we recall the following example in 
\cite{Ruo97}. 
Ruohonen discusses the event detection
problem: given a differential equation $y^\prime=f(t,y)$, with 
initial value $y(0)$, decide if a given condition
$g_j(t,y(t),y^\prime(t))=0, j=1,\cdots,k$ happens at some time $t$ in a
given interval $I$. Given the Turing machine
$\mathcal{M}$, the sequence $f_0,f_1,\cdots$ of rationals defined by
$$f_n = \left\{ \begin{array}{lll}
    2^{-m} & \textrm{if} & \textrm{$\mathcal{M}$ stops in $m$ steps on 
input $n$} \\
    0 & \textrm{if} & \textrm{$\mathcal{M}$ does not stop  on input $n$} 
\\
\end{array}
\right.$$ is not a computable sequence of rationals, but is a
computable sequence of reals, following the nomenclature of
\cite{PourEl:comap}. 
Now, the detection of the event $y(t)=0$ for the ordinary differential equation $y^\prime=0$, given $n$, and the initial value $y(0)=f_n$, is undecidable over any interval containing $0$, because $f_n=0$ is undecidable.  

A further modification can be obtained as follows \cite{Ruo97}. 
He defines 
the smooth function $$g(x)=f_{\lfloor x+1/2 \rfloor} e^{-\tan^2 \pi x},$$
which  
is computable on $[0,\infty)$. The detection of the event
$y_1(t)=0$ for the ODE
$$\left\{ \begin{array}{lll}
 y^\prime_1 &=& g(y_2)-1 \\
 y^\prime_2 &=& 0 \\
\end{array} \right.$$
given an initial value $y_1(0)=1$, $y_2(0)=n$, where $n$ is a nonnegative integer is then undecidable on $[0,1]$.

As put forth in \cite{Asa95} undecidability results given by recursive analysis are somehow built similarly.

\subsection{Dynamic undecidability}

To be able to discuss in more detail computability of differential
equations, we will focus on ODEs that encode the transitions of a
Turing machine instead of the result of the whole computation
simulation\footnote{This is called dynamic 
undecidability in
  \cite{Ruo93}.}.
Typically, we start with some (simple) 
computable injective function which encodes any 
configuration of a 
Turing machine $M$ as a point in $\R^n$. Let $x$ be the encoding of the 
initial configuration of $\mathcal{M}$. Then, we look for a function $f: 
E \subset \R^{n+1} \to \R^n$ such that the solution of 
$y^\prime(t)=f(y,t)$, with $y(0)=x$, at time $T \in N$ is the 
encoding of the configuration of $\mathcal{M}$ after $T$ steps.  We will 
see, in the remainder of this section, that $f$ can be restricted to 
have low dimension, to be smooth or even analytic, or to be defined on a 
compact domain.

Instead of stating that the property above is a Turing machine simulation, we 
can address it as a reachability result. Given the IVP defined by $f$ 
and $x$, and any region $A \subset\R^n$, we are interested in 
deciding if there is a $t \geq 0$ such $y(t) \in A$, i.e., if the flow 
starting in $x$ crosses $A$. It is clear that if $f$ simulates a Turing 
machine in the previous sense, then reachability for that system is 
undecidable (just consider $A$ as encoding the halting configurations of 
$\mathcal{M}$).
So, reachability is another way to address the computability of ODEs and 
a negative result is often a byproduct of the simulation of Turing machines. 
Similarly,  undecidability of 
event detection follows from Turing simulation results.

Computability of reachable and invariant sets have been investigated
in  \cite{Col05}  for continuous
  time systems and in \cite{Collins05} for  hybrid systems.


  In general, viewing Turing machines as dynamical systems provides them
  a physical interpretation which is not provided by the von Neumann
  picture \cite{Campagnolo}.  This also shows that many qualitative
  features of (analog or non-analog) dynamical systems, e.g. questions
  about basins of attraction, chaotic behavior or even periodicity,
  are non computable \cite{Moo90}.  Conversely, this brings into the
  realm of Turing machines and computability in general questions
  traditionally related to dynamical systems. These include in
  particular the relations between universality and chaos
  \cite{Asa95}, necessary conditions for universality
  \cite{DelKurBlo04}, the computability of entropy \cite{Koi01},
  understanding of edge of chaos \cite{LM05}, and relations with the
  shadowing property \cite{Hoyrup}.



\subsection{Embedding Turing machines in continuous time}



The embedding of Turing machines in continuous dynamical systems is 
often realized in two steps. Turing machines
are first embedded into analog space discrete time systems, and then the
obtained systems are in turn embedded into analog space and time
systems. 

The first step can be realized with low dimensional systems with 
simple dynamics: \cite{Moo90}, \cite{Ruo93}, \cite{Bra95}, 
\cite{Ruo97} consider general dynamical systems, \cite{KCG94}
piecewise affine maps, \cite{SS95} sigmoidal neural nets, \cite{KM97}
closed form analytic maps, which can be extended to be robust 
\cite{dsg05}, and
\cite{Potapov05} one dimensional very restricted piecewise
defined maps.

\newcommand{\figdir}{.}

\begin{figure}[h]
\begin{tabular}{ll}
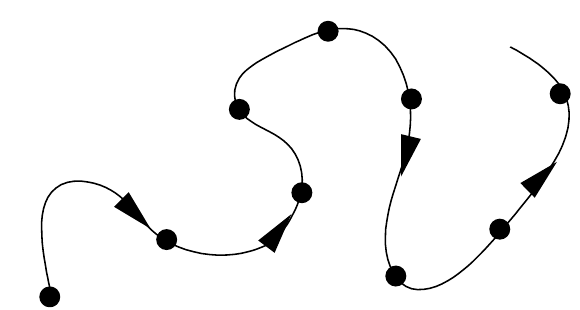
&
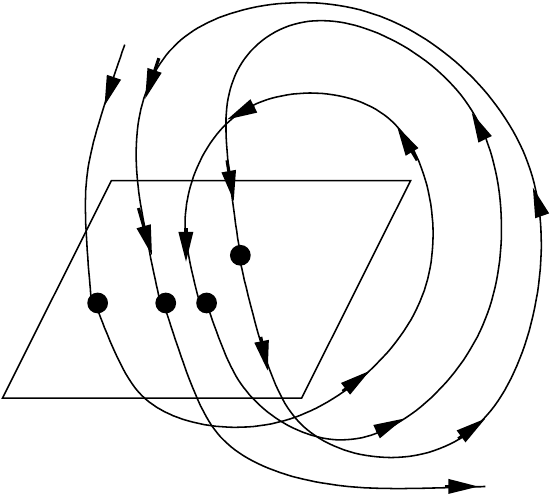
\\ 
\end{tabular}
\caption{Stroboscopic map (on left) and Poincaré map (on right) of the
  dynamic of a continuous time system.}
\label{fig:poinca}
\end{figure}

For the second step, the most common technique is to build a continuous time
and space system whose discretization corresponds to the embedded
analog space discrete time system. 

There are several classical ways to
discretize a continuous time and space system: see Figure~\ref{fig:poinca}. 
One way is to use a virtual stroboscope: the flow
$x_t=\phi(t,x)$, when $t$ is restricted to integers, defines the
trajectories of a discrete time dynamical system.  Another possibility
is through a Poincaré section: the sequence $x_t$ of the intersections
of trajectories with, for example, a hypersurface can provide the flow of a
discrete time dynamical system. See \cite{HSD03}.  

The opposite operation, called \textit{suspension}, is usually
achieved by extending and smoothing equations, and usually requires
higher dimensional systems.  This explains why Turing machines are
simulated by three-dimensional smooth continuous time systems in
\cite{Moo90}, \cite{Moo91}, \cite{Bra95} or by three-dimensional
piecewise constant differential equations in \cite{AMP95}, while they
are known to be simulated in discrete time by only two-dimensional
piecewise affine maps in \cite{KCG94}. It is known that
two-dimensional piecewise constant differential equations
cannot\footnote{See also already mentioned generalizations of this
  result in \cite{CV95} and \cite{Asarin:2001:DRP}.}  simulate
arbitrary Turing machines \cite{AMP95}, while the question whether
one-dimensional piecewise affine maps can simulate arbitrary Turing
machines is open. Other simulations of Turing machines by continuous
time dynamical systems include the robust simulation with polynomial
ODEs in \cite{dsg05}, \cite{gcb}. This result is an improved version
of the simulation of Turing machines with real recursive functions in
\cite{mlc:moo:fgc:99a}, where it is shown that smooth but non-analytic
classes of real recursive functions are closed under iteration.
Notice that while the solution of a polynomial ODE is computable on
its maximal interval of existence (see
Section~\ref{section:computabilityODEs}), the simulation result shows
that the reachability problem is undecidable for polynomial ODEs.

In addition to Turing machines, other discrete models can be simulated by differential equations.  
Simulating two-counter machines can be achieved in two dimensions, or
even one dimension, at the cost of a discontinuous ODE \cite{Ruo97}. Simulating
cellular automata can be done with partial differential equations
defined with $C^\infty$ functions \cite{Omohundro84}.

Notice that reversible computations of Turing machines (or counter
machines, or register machines) can be simulated by ODEs with
backward-unique solutions \cite{Ruo93}.


Continuous time dynamical systems can 
in turn be
embedded into other continuous time systems. For example, \cite{MJS07}
proves that a large class $S_n$ of systems of differential equations
are universal for analog computing on time-varying inputs in the
following sense: a system of this class can reply to some external
input $u(t)$ with the dynamics of any n$^{th}$ order differential
equation of the form
$z^{(n)}(t)=G(z(t),z^\prime(t),\cdots,z^{(n-1)}(t))+u(t)$, if a
suitable memoryless feedback and readout functions are added. 
As the n$^{th}$ order differential equation above 
can simulate Turing machines, systems from $S_n$ have the
power of a universal Turing machine. But since $G$ is arbitrary,
systems from $S_n$ can actually simulate any conceivable continuous
dynamic response to an input stream. Moreover, this results holds for
the case where inputs and outputs are required to be bounded.

\subsection{Discussion issues}

The key technique in embedding the time evolution of a Turing machine in a 
flow is to use ``continuous clocks'' as in  
\cite{Bra95}.\footnote{Branicky attributes the idea of a
  two phase computation to \cite{Bro89} and \cite{bro88}. A similar
  trick is actually present in \cite{Ruo93}. We will actually not
  follow \cite{Bra95} but its presentation in \cite{Campagnolo}.}  

The idea is to start from the function $f: \R \to \R$, preserving the 
integers, and build the
ordinary differential equation over $\R^3$ 
$$
\begin{array}{lll} 
y_1' &=  & c(f(r(y_2))-y_1)^3 \theta(\sin (2\pi y_3)) \\
y_2' &= & c(r(y_1) - y_2)^3 \theta(-\sin(2\pi y_3)) \\
y_3' & = & 1. \\
\end{array}
$$
Here $r(x)$ is a rounding-like function that has value $n$ whenever
 $x \in [n-1/4,n+1/4]$ 
for some integer $n$, and 
$\theta(x)$ is $0$ 
for $x \leq 0$, $\exp(-1/x)$ for $x>0$, and $c$ is some suitable 
constant.

The variable $y_3=t$ is the time variable. 
Suppose $y_1(0)=y_2(0)=x \in \N$. For $t \in [0,1/2]$,
$y_2'=0$, and hence $y_2$ is kept fixed to $x$. Now, if $f(x)=x$, then $y_1$ will be kept to $x$. If $f(x) \neq x$, then
$y_1(t)$ will approach $f(x)$ on this time interval, and from the
computations in \cite{Campagnolo}, 
if a large enough number is chosen for $c$ we can be sure that
$|y_1(1/2) - f(x) | \leq 1/4$. Consequently, we will have
$r(y_1(1/2))=f(x)$.  Now, for $t \in [1/2,1]$, roles are inverted:
$y_1'=0$, and hence $y_1$ is kept fixed to the value $f(x)$.  On that
interval, $y_2$ approaches $f(x)$, and $r(y_2(1))=f(x)$. The equation
has a similar behavior for all subsequent intervals of the form
$[n,n+1/2]$ and $[n+1/2,n+1]$.  Hence, at all integer time $t$,
$f^{[t]}(x) = r(y_1(t))$.\footnote{$f^{[t]}(x)$ denotes de $t$th
  iteration of $f$ on $x$.}  \cite{LCM07} proposes a similar
construction that returns $f^{[\lfloor t \rfloor]}(x)$ for all $t \in
\R$.

In other words, the construction above transforms a function over $\R$
into a higher dimensional ordinary differential equation that
simulates its iterations. To do so, $\theta(\sin (2\pi y_3))$ is used as
a kind of clock. Therefore, the construction is essentially ``hybrid''
since it combines smooth dynamics with non-differentiable, or at least
non-analytic clocks to simulate the discrete dynamics of a Turing
machine.
Even if the flow is smooth (i.e. in $C^\infty$) with 
respect to time, the orbit does not admit a tangent at every point since 
$y_1$ and $y_2$ are alternatively constant.
Arguably, one can 
overcome this limitation by restricting Turing machine simulations to 
analytic flows and maps. 
While it was shown that analytic maps over 
unbounded domains are able to simulate the transition function of any 
Turing machine in \cite{KM97}, only recently 
it was shown that Turing machines can be simulated with analytic flows over unbounded domains in \cite{dsg05}. It would be 
desirable to extend the result to compact
domains. However, it is conjectured in \cite{Moor98} that this is not 
possible, i.e. 
that no analytic map on a compact finite-dimensional space can
simulate a Turing machine through a reasonable input and output
encoding.

\subsection{Time and space contractions}

Turing machines can be simulated by ODEs in real time: for example, in the constructions we described above,
the state $y(T)$ at time $T \in N$ of the solution of the ordinary
differential equation 
encodes the state after $T$ steps of the Turing machine.
 However, since continuous time systems might undergo
  arbitrary space and time contractions, Turing machines, as well as
  accelerating 
 Turing machines\footnote{Similar possibilities of simulating accelerating 
Turing machines through quantum mechanics are discussed in \cite{CP01}.} \cite{Davies01}, \cite{Copeland98}, \cite{Cop02} or
  even oracle Turing machines, can actually be simulated in an arbitrary
  short time. 

In the paragraphs below, we will follow Ruohonen \cite{Ruo93} who  
denotes a continuous time system  by the triplet $(F,n,A)$, where $F$
defines the ordinary differential equation $y^\prime=F(y)$ over
$\R^n$, with accepting set $A$: some input $x$ is accepted iff the
trajectory starting with initial condition $x$ crosses $A$.

A machine $\mathcal{M}=(F,n,A)$  can be accelerated: the substitution $t=e^u -1$ 
for instance changes  $\mathcal{M}$ to
$((G,1),n+1,A\times \R)$ where
$$\frac{dg}{du} = G(g(u),u)= F(g(u))e^u \mbox{ and } g(u) = y(e^u -1),$$ 
yielding an exponential time acceleration. Note that the derivatives of 
the solution with respect to the new time variable $u$ are exponentially 
larger. 
Furthermore, the substitution $t= \tan( \pi u /2)$ gives an infinite time acceleration,
i.e. compresses any computation, even an infinite one, into the finite
interval $0 \le u <1$. Now, the derivatives go to infinity during the 
course of computation.

Turning to space contraction, replacing the state $y(t)$ of the machine $\mathcal{M}=(F,n,A)$
by $r(t)=y(t)e^{-t}$ gives an exponentially downscaled machine
$((H,1),m+1,H_1)$ where
$$ \frac{dr}{dt} = H(r(t),t) = F(r(t) e^t ) e^{-t} - r(t)$$
and
$$H_1= \{(e^{-t} q,t) | q \in A \mbox{ and } t \ge 0\}.$$ 
Obviously, this transformation reduces exponentially the distance between 
trajectories, which require increased precision to be distinguished.

Hardness results in the levels of the arithmetical or analytical
hierarchy for several decision problems about continuous time systems
are derived from similar constructions in \cite{Ruo93},
\cite{Ruo94}, \cite{Moo95b}, and \cite{AM95}.  Completeness results, 
as well as exact characterizations of the recognition power of piecewise
constant derivative systems according to their dimensions have been
obtained in \cite{Bou97b} and  \cite{These}.
Notice that such phenomena are instances of the so-called
Zeno's phenomena in hybrid systems literature: \cite{AD91} and \cite{AM95}.

It can be observed that previous constructions yield undecidability
results only for functions over infinite or half-open intervals, since
positive reals, corresponding to Turing machines integer time, are
mapped to intervals of the form $[0,1)$. An analytical construction
is indeed possible over a finite closed domain of the form $[0,1]$,
with a function $G$ which is continuous and bounded on $[0,1]$, but
non-differentiable. It follows that the event detection problem, for
example, is undecidable even with continuous functions over compact
intervals \cite{Ruo97b}. 

Undecidability is ruled out, 
however, if the function $G$ is sufficiently smooth (say, in  
$\mathcal{C}^1$), if both $G$ and the initial value are computable, and 
if a
sufficiently robust acceptance condition is considered. 
Indeed, problems such as the event
detection problem then become decidable, since 
the system
can be 
simulated effectively \cite{Ruo97b}.

Instead of embedding Turing machines into continuous dynamical
systems, it is natural to ask 
if there is a better way to think about 
computation and complexity for the dynamical
systems that are commonly used to model the physical world.  We address 
this issue in the next section.

\section{Toward a complexity theory}
\label{sec:quatre}

Here we discuss a number of different views on the complexity of
continuous dynamical systems. We consider general systems and question the difficulty of simulating 
continuous time systems with a digital
model. 
We then focus on dissipative systems, where trajectories converge to 
attractors. In particular, we discuss the idea that the 
computation time should be the natural time variable of the ODE. 
Finally, we review complexity results for more general continuous time systems that correspond to classes of real recursive functions.

\subsection{General continuous dynamical systems}

In \cite{VSD86} it was asked if analog computers can be more efficient
than digital ones. Vergis {\em et al.} also postulated the `Strong Church's Thesis'', which states that the time required by a digital computer to simulate any analog
computer is bounded by a polynomial function of the resources used by
the analog computer. They claim that
the Strong Church's Thesis is provably true for continuous time
dynamical systems described by any Lipschitzian ODE $y^\prime=f(y)$.

The resources used by an analog computer include the time interval of operation, 
say $[0,T]$, the size of the system, which can measured by $\max_{t \in 
[0,T]} ||y(t)||$, as well as the bound on the derivatives of $y$. For 
instance, mass, time of operation, maximum displacement, velocity, 
acceleration and applied force are all resources used by a particle 
described by Newtonian mechanics \cite{VSD86}. 

The claim above depends on the definition of  ``simulation''.  In the article
\cite{VSD86} it is considered that the  IVP $y^\prime=f(y)$, 
$y(0)=x$ is simulated if, given T and some precision $\varepsilon$, one 
can 
compute an approximation of $y(T)$ with a margin of error of at most $\varepsilon$. 
Using Euler's method to solve this problem, and considering that the 
round-off error is less than $\sigma$, the total error bound is given 
by 
\begin{equation}
||y(T)-y^{*}_N|| \le \frac{h}{\lambda} 
\left[\frac{R}{2}+\frac{\sigma}{h2}\right](e^T \lambda -1),
\label{vsd:bound}
\end{equation}
where $y^{*}_N$ is the approximation after $N$ steps, $h$ is the step 
size, $\lambda$ is the Lipschitz constant for $f$ on $[0,T]$, and 
$R=\max\{||y^{\prime\prime}(t)||, t \in [0,T]\}$. From the bound in 
(\ref{vsd:bound}), Vergis {\em et al.} conclude that the number $N$ of 
necessary steps in Euler's method is polynomial in $R$ and 
$\frac{1}{\varepsilon}$. They use this fact to claim that the Strong 
Church's Thesis is valid for ODEs. However, $N$ is exponential in $T$, 
which is the time of operation of this analog computer. This makes the 
argument in \cite{VSD86} inconclusive, as pointed out in \cite{Orponen}.

More recently, Smith discusses in \cite{Smi1} if hypercomputation is
possible with respect to the $n$-body problem in mechanics. In particular, he shows that the
exponential dependence in $T$ can be eliminated. As observed in
\cite{Smi1}, all classical numerical methods of fixed degree for
solving differential equations suffer from the same exponential
dependence in $T$. However, by considering a combination of
Runge-Kutta methods with degrees varying linearly with $T$, it is
possible to derive a method that only requires $N$ to be
polynomial in $T$, as long as the absolute value of each component of
$f$, $y$, and the absolute value of each partial derivative of $f$
with respect to any of its arguments, having total differentiation
degree $k$, is
in $(kT)^{\mathcal{O}(k)}$ \cite{Smi1}. The implications of these results for Strong
Church's Thesis are discussed in \cite{Smi1} and
\cite{Bou05}.

The same question can be addressed in the framework of recursive 
analysis. 
When $f: [0,1] \times [-1,1] \to
\R$ is polynomial time computable and satisfies a weak form of the
Lipschitz condition, the unique solution $y$ on $[0,1]$ of IVP
$y^\prime=f(t,y)$, $y(0)=0$ is always polynomial space computable
\cite{Ko83}.  Furthermore, solving in polynomial time a differential
equation with this weak Lipschitz condition is essentially as
difficult as solving a PSPACE-complete problem, since there exists 
a polynomial time computable function $f$ as above  
whose solution $y$ is not
polynomial time computable unless $P=PSPACE$ \cite{Ko83}, \cite{Ko91}.

Ko's results are not directly comparable to the polynomial bound shown 
in 
\cite{Smi1}. In recursive 
analysis, the input's size is the number of bits of precision. If the 
bound on 
the error of the approximation of $y(t)$ is measured in bits, i.e., if 
$\varepsilon=2^{-d}$, then the required number 
of steps $N$ in \cite{Smi1} 
is exponential in $d$. 

If $f$ is analytic, then the solution of $y^\prime=f(y)$ is also 
analytic. In that case, timestepping methods can be avoided. 
That is the approach followed in \cite{MM93}, where it is proved 
using recursive analysis that if $f$ is analytic and 
polynomial time computable then the solution is also polynomial time 
computable.

In short, although Strong Church's Thesis holds for analytic ODEs, it
has not yet been fully proved for general systems of Lipschitzian
ODEs. Hence, the possibility of super-polynomial computation by
differential equations cannot be ruled out, at least in principle. For
informal discussions on Strong Church's Thesis, refer
to \cite{Aar05} and \cite{McL01}.

Several authors have shown that certain decision or optimization
problems (e.g. graph connectivity, linear programming) can be solved
by specific continuous dynamical systems. Examples and references can be found in the papers 
\cite{Smi98}
\cite{VSD86}, 
\cite{bro88}, 
\cite{Ref6-BFFS03},  \cite{HM94a},  \cite{HSF02}



\subsection{Dissipative systems}

We now focus on dissipative systems and review two approaches. The
first is about neural network models, such as continuous Hopfield  
networks, that simulate circuits in a
non-uniform manner, and leads to lower bounds on the complexity of such
networks. The second deals with convergence to attractors, and  
considers  suitable energy functions as ways  
to measure the complexity of continuous time processes.

When considering dissipative systems, such as Hopfield neural
networks, the following approach to a complexity 
theory is 
natural.  Consider families $(C_n)_{n \in N}$ 
of continuous time
systems, for each input length $n \geq 0$.
Given some digital input $w \in \{0,1\}^*$, the
system $C_n$ evolves on  input $w$ (or some encoding of $w$), where $n$
is the length of $w$. It will eventually reach some stable
state, which is considered the result of the computation.

This circuit inspired notion of computability is the most common in the literature about the computational complexity
of neural networks models: see 
survey
\cite{SOSurvey}.
With respect to this approach, continuous time symmetric Hopfield
networks with a saturated linear activation function have been proved
to simulate arbitrary discrete-time binary recurrent neural networks,
at the cost of only a linear size overhead \cite{SO2000},
\cite{SO2003}.
This might be thought counterintuitive, since such symmetric networks,
which are constrained by some Liapunov energy function, can only
exhibit convergence phenomena, and hence cannot even realize a simple
alternating bit.  However, the convergence of dissipative  systems
can be exponentially long in the size of the system \cite{SimOrp03b},
and hence the simulation can be accomplished using a subnetwork that provides
$2^n$ clock pulses before converging.  

The languages recognized by polynomial size families of
discrete-time Hopfield networks have been proved in
\cite{nc:Orponen:1996} to correspond to non-uniform complexity class
$PSPACE/poly$ for arbitrary interconnection weights, and to $P/poly$
for polynomially bounded weights. Therefore, 
families of continuous time symmetric Hopfield networks have the same lower bounds. 
However, these lower bounds may be not tight, since upper bounds
for continuous time dynamics are not known \cite{SO2003},
\cite{SOSurvey}.

Let us now turn our attention to dissipative systems with a Lyapunov
function $E$. 

Gori and Meer \cite{GM02} consider a computational model
which has the capability of finding the minimizers (i.e. the points of
local or global minimum) of $E$. To prevent the complexity of a
problem from being hidden in the description of $E$, this function must be
easy to compute. In that setting, a problem $\Pi$ is considered easy if there
exists a unimodal function $E$ (i.e. all local minimizers of $E$ are
global minimizers) such that the solution of $\Pi$ can be obtained
from the global minimum of $E$.




More precisely,
Gori and Meer
investigate in \cite{GM02} a model where a problem $\Pi$ over the  
reals is considered to be solved
if there exist a  family $(E_n)_n: \R^n \times \R^{q(n)} \to \R$ of  
energy functions, given by a uniform family of straight line programs  
($q$ is some fixed polynomial),  and another family $(N_n)_n$ of  
straight line programs, such that
for all input $d$, a solution $\Pi(d)$ of the problem can be computed  
using $N_{q(n)}(w^*)$, from
a global minimizer $w^*$ of $w \to E_n(d,w)$.

Gori and Meer define classes $U$ and $NU$ in analogy with $P$ and $NP$
in classical complexity. $U$ corresponds to the above mentioned case
where  for all $d$, $w \to E_n(d,w)$ is unimodal, in opposition to $NU 
$ where it needs not be unimodal.
%
%
%
Notions of reductions are introduced, and it is
proved that the natural optimization problem ``find the minimum of some
linear objective function over a set defined by quadratic multivariate
polynomial constraints'' is $NU$-hard. They show that there exist
(artificial) $NU$ complete problems. These ideas are generalized to
obtain a polynomial hierarchy, with complete problems \cite{GM02}.


Actually, Gori and Meer's proposed framework is rather abstract, avoiding several
problems connected to what one might  expect of a true complexity theory
for continuous time computations. Nonetheless, it has the great advantage of
not relying on any particular complexity measure for the computation of 
trajectories. 
See the interesting
discussion in \cite{GM02}.

However, one would like 
to understand the complexity of approaching
the minima of energy functions, which correspond to the equilibria of 
dynamical systems.
First steps toward this end have been investigated in  \cite{HSF02}, where dissipative systems with exponential 
convergence are explored. 
Recall that if $x^*$ is a sink, then 
the rate of convergence toward $x^*$ 
satisfies $$|x(t)-x^* | \equiv e^{-\lambda t}$$ where $-\lambda$ is 
the largest
real part of an eigenvalue of $Df(x^*)$. 
This means that
$\tau=1/\lambda$ is a natural characteristic time of the attractor:
every $\tau \log 2$ time units, a new bit of the attractor is
computed.

For the systems considered in \cite{HSF02}, each sink has an attracting 
region, where the trajectories are trapped.
One can define the computation time $t_c$ of a dissipative
continuous time dynamical system as $t_c=max(t_c(\epsilon),t_c(U))$,
where $t_c(\epsilon)$ is the time required to reach some $\epsilon$
vicinity of some attractor, and $t_c(U)$ is the time required to reach
its attracting region. Then,
$T=\frac{t_c}{\tau}$ is a dimensionless complexity measure, invariant
under any linear time contraction.

Two continuous time algorithms, $MAX$ to compute the maximum of $n$
  numbers, and $FLOW$ to compute the maximum flow problem have been
  studied in this framework in \cite{HSF02}. $MAX$ has been shown to
  belong to proposed complexity class $CLOG$ (continuous log time),
  and $FLOW$ to $CP$ (continuous polynomial time). The authors
  conjecture that $CP$ corresponds to classical polynomial time
  \cite{HSF02}.
  Both $MAX$ and $FLOW$ algorithms are special cases of a flow
  proposed in \cite{F65} to solve linear programming problems, further
  investigated in \cite{BFFS03} and \cite{BFFS04}.  Variations on
  definitions of complexity classes, as well as
  ways to introduce non-deterministic classes in relation to
  computations by chaotic attractors have also been discussed in 
\cite{SS98}.


\subsection{Complexity and real recursive functions}

Real recursive functions are a convenient way to analyse the
computational power of certain operations over real
functions. Additionally, given a continous time model, if its
equivalence with a function algebra of real recursive functions can be
established, then some properties of the model can be proved
inductively over the function algebra.  

Since many time and space complexity classes have recursive
characterizations over $\N$ \cite{Clo95}, structural complexity
results about discrete operations may imply lower and upper bounds on the
computational complexity of real recursive functions. This approach
was followed in \cite{cam:moo:fgc:00} to show that $\relem$ contains
extensions of the elementary functions, and further developed  in  \cite{cam:tcs:04} to
obtain weaker classes that relate to the exponential space
hierarchy. This tells us something about the computational complexity
of certain dynamical systems. For instance, $\relem$ corresponds to
cascades of finite depth, each level of which depends linearly on its
own variables and the output of the level before it.

Results about the idea of lifting computability questions over $\N$ to
$\R$ have been discussed before. Concerning complexity, the question
$P=NP$ in classical complexity has been investigated using real
recursive functions by Costa and Mycka.  In particular, they propose
two classes of real recursive functions such that their inequality
would imply $P \ne NP$ in \cite{CM06c}, \cite{MC06}. More generally a part of
Costa and Mycka's program, explicitly stated in \cite{MC05Eatcs} and \cite{CM06b}, 
uses   recursion theory on the reals to establish a bridge between
computability and complexity theory and mathematical analysis.

\section{Noise and Robustness}
\label{sec:cinq}

Up to this point we have considered continuous time computations in
idealized, noise-free spaces.
As was also the case in the survey by Orponen \cite{Orponen}, most of
the results we discussed disregard the impact of noise and imprecision
in continuous time systems. This is a recurrently criticized weakness of
the research in the field. While there have not been major breakthroughs
with regards to these problems as they relate specifically to continuous
time computations, some interesting developments  concerning noise and
imprecision  have come about in discrete time analog computation
studies. In this section we will
broaden our scope to discuss a number of discrete time results. We
believe that some of these studies and results might be generalized to,
or at least provide some insight into, the effects of noise and
imprecision on continuous time systems, although this work has yet to be
done.


We first focus on systems with a bounded state space about which
a folklore conjecture
claims that robustness
implies decidability. We review some results that support this
conjecture as well as
others
that challenge it.
At the end of this section we discuss continuous time systems with
unbounded state spaces.


Common techniques to simulate Turing machines by dynamical systems in
bounded state spaces require the encoding of the configuration of the
Turing machine into real numbers. Since Turing machines have unbounded tapes (otherwise they would degenerate into finite automata), these simulations are destroyed if
the real numbers or the functions involved are not represented with
infinite precision. 
This leads to
%
the folklore conjecture,  popular in particular in the verification
community,
which states that undecidability do not hold for ``realistic'',
``unprecise'', ``noisy'', ``fuzzy'', or ``robust'' systems. See for
example \cite{Fra}, \cite{Foy04} for various statements of this
conjecture, and \cite{TalkEugene} for discussions on other arguments that
lead to this conjecture.


There is no consensus on what a realistic noise model is. A discussion
of this subject would require to question what are good models of the
physical world. In the absence of a generally accepted noise model, one
can however consider various models for noise, imprecision or smoothness
conditions, and investigate the properties of the resulting systems.
%

In particular, there have been several attempts to show  
that noisy
analog systems are
at best equivalent to finite automata.
Brockett proved that continuous time dynamical systems
can simulate arbitrary finite automata  in \cite{Bro89}. Using topological arguments
based on homotopy equivalence relations and associated Deck
transformations, he showed in \cite{Bro94} that some automata can be
associated to dissipative continuous time systems.

Maass and Orponen proved that
the presence of bounded noise reduces the power of a large set of
discrete time analog models to that of finite automata  in \cite{MO97}. This
extends
a previous result established in \cite{Cas96}, \cite{Cas98} for the
case where the output is assumed to be perfectly reliable (i.e. $\rho=1/2$ in what follows).

Maass and Orponen's idea is to replace a perfect discrete time dynamic
of type $x_{i+1}=f(x_i,a_i)$, where $a_i$ is the symbol input at time
$i$, over a compact domain, by a probabilistic dynamic
\begin{equation} \label{eq:etoile}
Probability(x_{i+1} \in B) = \int_{q \in B} z(f(x_i,a_i),q) d\mu,
\end{equation}
where
$B$ is any Borel set. Here, $z$ is a density kernel reflecting
arbitrary noise, which is assumed to be piecewise equicontinuous. This means that,  for
all $\epsilon$, there exists $\delta$ such that for all $r,p,q$,
$\|p-q\| \leq \delta$
implies $|z(r,p) - z(r,q)| \leq \epsilon$. They denote by $\pi_x(q)$ the
distribution of states after string $x$ is
processed from some fixed initial state $q$, and
they consider the following robust acceptance
condition: a language $L$ is recognized, if there exists
$\rho>0$ such that $x \in L$ iff $\int_F \pi_{xu}(q) d\mu \geq 1/2+
\rho$ for
some $u \in \{U\} ^*$, and $x \not\in L$ iff $\int_F \pi_{xu}(q) d\mu
\le
1/2- \rho$ for all $u \in \{U\} ^*$, where $U$ is the blank symbol, and
$F$ the set of accepting states.
Then, they show that the space of functions
$\pi_{x}(.)$  can be partitioned into finitely
many classes $C$ such that two functions $\pi_x(.)$ and $\pi_y(.)$ in
the same class satisfy $\int_r |\pi_x(r)-\pi_y(r)| d\mu \leq \rho$. Therefore, two
words $x,y$ in the same class satisfy $x w \in L$ iff $y w \in L$ for
all words $w$.

In fact, for any common noise, such as Gaussian noise, which is nonzero
on a sufficiently large part of the state space, 
systems described by (\ref{eq:etoile}) are
unable to recognize even arbitrary regular languages \cite{MS99}.
They recognize precisely the definite languages introduced by
\cite{Rabin63}, as shown in \cite{MS99} and 
\cite{Ben-Hur:2004:PAA}. If the noise level is bounded, then
all regular languages can be recognized \cite{MO97}. 
Feedback continuous time circuits in \cite{MJS07} have the same computational power when subject to bounded noise.

As an alternative to the probabilistic approach of Maass and Orponen, noise can be modeled through non-determinism.  One can associate to
deterministic noise-free discrete time dynamical system $S$ defined by
$x_{i+1}=f(x_i)$,  the non-deterministic $\epsilon$-perturbated
system $S_{\epsilon}$ whose trajectories are sequences $(x_n)_n$ with
$\|x_{i+1}-f(x_i)\| \leq \epsilon$. For a dynamical system $S$, it is
natural to consider the predicate $Reach[S](x,y)$ (respectively
$Reach_n[S](x,y)$), which is true if there exists a trajectory of $S$
from $x$ to $y$
(resp. in $i \leq n$ steps).
Then, algorithmic verification of safety of state properties is closely
related to the problem of computing reachable states. Given $S$, and 
a subset of initial states $S_0$, let $Reach[S]$ denote the set of
$y$'s such that $Reach[S](x,y)$ for some $x \in S_0$. Given a state  
property $p$ (i.e. a property which is either true, or false in a  
state $s$), let $[[\neg p]]$ denote the subset of states $s$ where $p 
$ is false. Then $S$ is safe ($p$ is an invariant) iff
$Reach[S] \cap [[\neg p]]  = \emptyset$
(see, for  example, \cite{ACH+95} and \cite{NOSY93}).

If the class of systems under consideration is such that relation
$Reach_n[S](x,y)$ is recursive\footnote{Recursive in $x,y$ and
   $n$.} (assuming that $S_0$  recursively enumerable), then $Reach[S]$ is  
recursively enumerable because $Reach[S] =
\bigcup_n
Reach_n[S]$. Several papers have been devoted to prove that $Reach[S]$
is indeed recursive for classes of dynamical systems
under different notions of robustness. We now review several of them.

Fränzle observes in \cite{Fra} that the computation of
$Reach[S_\epsilon]$
by $Reach[S_\epsilon ] = \bigcup_n Reach_n[S_\epsilon]$ must always
terminate if
$Reach[S_\epsilon]$ has a strongly finite diameter. This means that
there exists an
infinite number of points in $Reach[S_\epsilon]$ at mutual distance of
at least
$\epsilon$, which is not possible over a bounded domain.
It
follows that if we call robust  a system which is either unsafe, or
whose $\epsilon$-perturbated system is safe for some $\epsilon$, then
safety is
decidable for robust systems over compact domains \cite{Fra}.

Consider as in \cite{P98} the relation $Reach_\omega[S]=
\bigcap_{\epsilon >0} Reach
[S_\epsilon]$, corresponding to states that stay reachable when noise
converges to $0$. Asarin and Bouajjani prove in
\cite{asarin01perturbed} that for large classes of discrete and
continuous time dynamical systems (Turing machines, piecewise affine
maps, piecewise constant differential equations), $Reach_\omega[S]$ is
co-recursively enumerable.  Furthermore, any co-recursively enumerable
relation is of form $Reach_\omega[S]$ for some $S$ for the classes that Asarin and Bouajjani consider. 
Therefore, if we call robust a system
such that $Reach[S]=Reach_\omega[S]$, then computing $Reach[S]$ is
decidable for robust systems.

Asarin and Collins considered in \cite{AsaCol05} a model of Turing
machines exposed to a small stochastic noise, whose computational
power have been characterized to correspond to $\Pi^0_2$. It is
interesting to compare this result with previous results where a small
non-deterministic noise lead to $\Pi^0_1$ (co-recursively enumerable
sets) computational power only.

We now turn our attention to results that challenge the conjecture
that robustness implies decidability.  A first example is that  
the safety of a system is 
still undecidable
if the transition relation of the system is open,
as proved in \cite{HR00}, and \cite{TalkEugene}. However,
the question for the restriction to a uniform non-deterministic noise  
bounded from below is open
\cite{TalkEugene}.

Noise can also be modeled by perturbating trajectories. Gupta,
Henzinger and Jagadeesan consider in \cite{GHJ97} a metric
over trajectories of timed automata, and assume that if a system
accepts a trajectory, then it must accept neighboring trajectories also.
They prove that this notion of robustness is not sufficient to avoid
undecidability of complementation for Timed automata. Henzinger and
Raskin prove in \cite{HR00} that major undecidability results about
verification of hybrid systems are still undecidable for robust
systems in that sense.

Finally, we review a recent robustness result for continuous time
dynamical systems with unbounded state space.
Graça, Campagnolo and Buescu prove in \cite{gcb}
that polynomial differential equations can simulate robustly Turing
machines in real time. More precisely, let us consider that $\theta :
\N^3 \to \N^3$ is the transition function some Turing machine $M$ whose
configuration is encoded on $\N^3$. Then, there is a $\epsilon >0$, a
solution $f$ of a polynomial ODE, and an initial condition $f(0)$ such that the solution
of $y^\prime=f(t,y)$
encodes the state of $M$ after $t$ steps with error at most $\epsilon$.
Moreover, this holds for a neighborhood of any integer $t$ even if $f$ and the initial
condition $f(0)$ are perturbed.  Obviously, this kind of simulation
requires the system to have an unbounded  state space.




\section{Conclusion}
\label{sec:six}

Having surveyed the field of
continuous time computation theory, we see that it provides insights into many diverse areas  
such as
verification, control theory, VLSI design, neural networks, analog
machines, recursion theory, theory of differential equations and computational complexity.

We have attempted to give a systematic overview of the most relevant models and results on continuous time computations. In the last decade many new results have been  obtained, indicating that this is an active field of research. We reviewed recent developments of the theory of continuous time computation with respect to computability, complexity and robustness to noise, and we identified several open problems. To conclude, we will discuss some directions for future research related to these areas.

\subsubsection{Computability.} It is not clear if a unifying concept similar to  the Church-Turing thesis exists for continuous time computation. Although it has been shown that some continuous time models exhibit super Turing power, these results rely on the use of an infinite amount of resources such as time, space, precision, or energy. In general, it is believed that ``reasonable'' continuous time models cannot compute beyond Turing machines. This raises the question if physically realistic continuous time computation can be  as powerful as digital computation. We saw that if we restrict continuous time systems to evolve in a bounded state space and to be  subjected to noise, then they become comparable to finite automata. However, with a bounded state space, Turing machines also degenerate into
finite automata. Since analytic and robust continuous time systems can
simulate Turing machines in an unbounded state space, we believe that
digital computation and analog continuous time computation are equally
powerful from the computability point of view.  Moreover, as we saw, several recent results establish the equivalence between  functions computable by polynomial ODEs, GPAC-computable functions and real computable functions in the framework of recursive analysis. These kind of results reinforce the idea that there could be an unified framework for continuous time computations, analogous to what occurs in classical computation theory.

We feel that a general paradigm of realistic continuous time computations ideally should only involve analytic functions, since these are often considered as the most acceptable from a physical point of view. Continuous dynamical systems are a natural form of representing continuous time processes. Classical systems like the van der Pol equation, the Lotka-Volterra system or the Lorenz equations are described with differential equations with an analytic, even polynomial, right-hand side. These physics-related arguments combined with the computability properties of systems of polynomial differential equations lead us to suggest that this continuous time model is a possible candidate for a general paradigm of continuous time computation. We believe that this idea deserves further investigation.


\subsubsection{Complexity.} We saw that a complexity theory for continuous time computation is still under way and that there has not been an agreement between authors on basic definitions such as computation time or input size. The results described in Section~\ref{sec:quatre} are either derived from concepts which are intrinsic to the continuous time systems under study or related to classical complexity theory. As computable analysis is a well established and understood  
framework for the study of computational complexity of continuous  
time systems, we believe that understanding relations between  
different approaches and computable analysis from a complexity point  
of view is of first importance. 
There are still many open questions about upper bounds for continuous time models. For example, upper bounds are not known for Hopfield networks and general systems of Lipschitizian ODEs, which compromises the validity of the Strong Turing thesis. We saw that this thesis might hold for systems of analytic ODEs. This leads us to ask  if a continuous time computation theory based on polynomial ODEs could be naturally extended to a complexity theory.

Computable analysis also permits to study the complexity of real recursive functions. One of the most intriguing area of research in continuous time computation tries to explore the link between real recursive functions and computational complexity to establish a translation of open problems of classical complexity into Analysis.

\subsubsection{Robustness.} We saw that very little research has been done with respect to the robustness and tolerance to noise of  continuous time systems. 
One might ask how the power  
of analog computations increases with their precision. This question  
was raised and formalized for discrete time analog systems, in particular for dynamical recognizers, in \cite 
{Moor98} but most of the research in that  
direction has yet to be done. Many interesting open questions arise  
if one asks if undecidability results for continuous time systems  
still hold for robust systems. This is of first importance for  
example for the verification of hybrid systems, since this question is  
closely related to the question of  termination of automatic  
verification procedures. A better understanding 
of the hypotheses under which noise 
yield 
decidability or undecidability  is required.  For  
example, non-deterministic noise on open systems does not  
rule out undecidability, but the question is unanswered for a uniform  
noise bounded from below \cite{TalkEugene}.



\bigskip
\noindent{Acknowledgments.} 
 We would like to thank all our colleagues in a wide sense, since this survey benefited from recent and old discussions with a long list of people. Many of them have their work cited in the text. 
We would also like to deeply thank
 Kathleen Merrill for her careful reading of the text and for her suggestions to improve its clarity, and an anonymous referee for his/her helpful advice.
 This work was partially supported by EGIDE and GRICES under the Program 
\emph{Pessoa} through the project \emph{Calculabilit\'{e} et
complexit\'{e} des mod\`{e}les de calculs \`{a} temps continu}, by 
\emph{Funda\c{c}\~{a}o para a Ci\^{e}ncia e a Tecnologia} and FEDER via the
Center for Logic and Computation - CLC and the project ConTComp
POCTI/MAT/45978/2002. This work was also supported by French Ministery
of Research through ANR Project SOGEA.

\findocument

\end{document}